\documentclass[12pt,a4paper]{article}
\usepackage[utf8]{inputenc}
\usepackage{amsmath}
\usepackage{amsfonts}
\usepackage{amssymb}
\usepackage{mathtools}
\usepackage{graphicx}
\usepackage{cite}
\usepackage[capitalise]{cleveref}
\graphicspath{ {./figures/} }
\usepackage[left=2cm,right=2cm,top=2cm,bottom=2cm]{geometry}
\crefformat{section}{#2section~#1#3}
\author{T.V. Obikhod, S.B. Chernyshenko}
\title{\bf{Searches for axion-like particles in proton-proton and ion-ion collisions at energies in the center of mass system of 5.02 TeV and 13 TeV}}
\date{%
    {\it Institute for Nuclear Research NAS of Ukraine, Kyiv 03028, Ukraine}\\%
    \today
}
\begin{document}

\maketitle

\section{Abstract}

	As part of experimental measurements by the ATLAS collaboration during 2015 and 2018 years of light-on-light scattering in Pb+Pb collisions, at an integrated luminosity of 2.2 nb$^{-1}$ and energy of 5.02 TeV, we modeled the cross sections of the axion-like particle (ALP) production, candidates for dark matter, with subsequent decay into gamma quanta. The obtained calculations showed the dependence of the cross section on the number of ALPs events, on the energy in the center of mass system in proton-proton and Pb-Pb collisions and on the mass region of the ALPs. Using four models with the  parameters, which specify the Good-Walker eigenstates, we calculated the ALP production cross sections  in proton-proton collisions for the single dissociation and diffractive dissociation components.

\section{Introduction}
New experimental data connected with the observation of the light-by-light scattering in Pb+Pb and proton-proton collisions are a new and promising area of both theoretical research and experimental measurements. Published data from the ATLAS collaboration reported the observation of light-by-light scattering with a significance of 8.2 standard deviations \cite{Ref1.}. Such light-by-light scattering may be connected with mediation of ALP studied by ATLAS Forward Proton spectrometer in proton-proton collisions in 2017 at a centre-of-mass energy of $\sqrt{s}$ = 13 TeV with selected 441 candidate for signal events \cite{2.}. Representative searches of BSM physics in ultraperipheral collisions of Pb+Pb are also connected with ALP particles \cite{3.}. Feynman diagrams for ALP particle productions are presented in Fig. 1
\begin{center}
\includegraphics[width=0.75\textwidth]{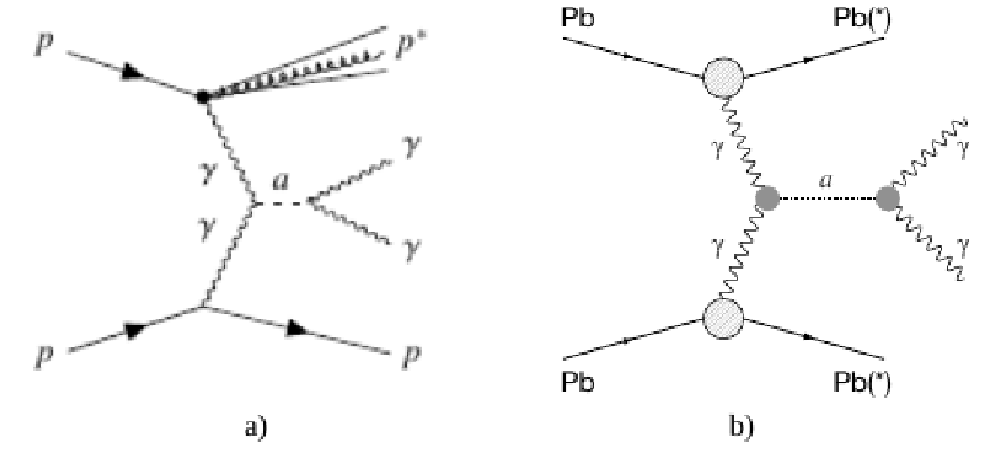}\\
\emph{\textbf{Fig.1}}{\emph{Feynman diagrams for ALP particle productions in a) proton-proton-collisions b) Pb-Pb collisions. }}
\end{center}
The search for ALP - dark matter candidates in mass ranges from several eV to 2 TeV and in various types of experimental installations, from space observatories to colliders, is an urgent and complex task. Therefore, carrying out computer modeling using a modern program SuperChic v4.2 \cite{4.} to clarify the details and features of the ultra-peripheral process is a component of our paper. Our article is devoted to modeling proton-proton and ion-ion (Pb+Pb) collisions, taking into account the peculiarities of the kinematics of the process under consideration and finding the optimal mode for ALP searching.

\section{Theory of ultra-peripheral nuclear collisions}
In ultraperipheral collisions (UPC) of accelerated nuclei, in which the impact parameter $b$ exceeds the sum of the radii of the nuclei 
$(b > R1 + R2)$, the interactions of nuclei can be described by the Weizsecker-Williams equivalent photon method, according to which the Lorentz-squeezed Coulomb fields of nuclei can be presented as a stream of photons. As for proton-proton interaction described by Good-Walker (GW) formalism, it is high energy diffractive interaction described in terms of the exchange of the pomeron, which may distort the wave function of the incoming proton, leading to the production of higher nucleon resonances. At high energies proton is formed by valence quarks, whose position in the impact parameter plane is fixed. The ideology of ion-ion and proton-proton ultraperipheral collisions connected with unchanged impact parameter coordinates, so the formula for the cross section for pomeron exchange between two dipoles is the following
\[\sigma_{ab}=\int\frac{dk^2_T}{k^4_T}\alpha^2_s[1-F_a(4k^2_T)][1-F_b(4k^2_T)],
\]
where $F_i(4k^2_T)$ are the form factors of the incoming colourless dipoles. As at small and at a larger $k_T$ the value of the cross section is specified by the cutoff induced by the pomeron, there is no dispersion and the interaction to destroy the coherence of the wave functions of the incoming protons. So, the probability of diffractive dissociation is negligible, that will be presented later by our calculations. Two-channel eikonal approximation and the parametrised form factor with $i, k = 1, 2$ of the form
\[F_i(t)=exp(-(b_i(c_i-t))^{d_i}+(b_ic_i)^{d_i}),\]
with the six parameters $b_i , c_i , d_i$ , is tuned to describe the elastic scattering data. Table 1 presented the  parameters of the four models of the two-channel proton-proton scattering data.\\
\begin{center}
Table 1 {\it The  parameter specification of the GW eigenstates}
\begin{tabular}{|c|c|c|c|c|} 
 \hline
 $|a|^2$ & 0.46 & 0.25 & 0.24 & 0.25 \\ 
 \hline
 $b_1 (GeV^{-2})$ & 8.5 & 8.0 & 5.3 & 7.2 \\
 \hline
 $c_1 (GeV^{2})$ & 0.18 &0.18& 0.32 &0.53 \\
 \hline
 ${d}_{1}$ & $0.45$ & $0.63$ & $0.55$& $0.6$ \\
 \hline
 $b_2 (GeV^{-2})$ & 4.5 & 6.0 & 3.8 & 4.2 \\
 \hline
 $c_2 (GeV^{2})$ & 0.58 & 0.58 & 0.18 &  0.24 \\
 \hline
 $d_2$ & 0.45 & 0.47 & 0.48 & 0.48 \\
 \hline
\end{tabular}
\end{center}
Eikonal absorptive corrections lead to the factor ${S}_{ik} = exp(-\Omega_{ik} /2)$, which describes the multiple rescattering, shown as $S_{ik}$ in Fig. 2. 
\begin{center}
\includegraphics[width=0.5\textwidth]{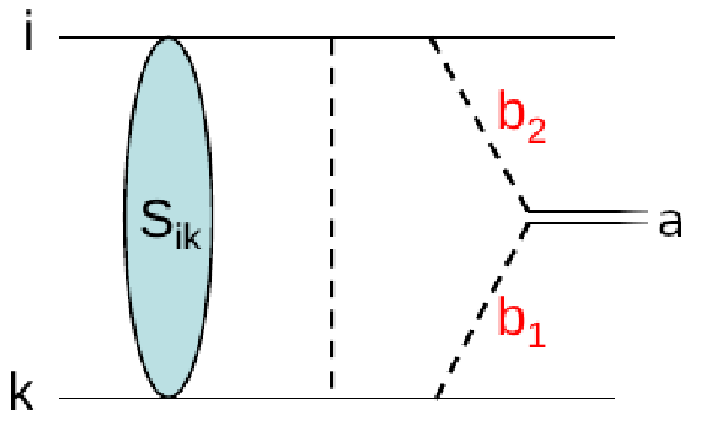}\\
\emph{\textbf{Fig.2}}{\emph{The diagram describing the amplitude of ALP production, with i, k initial states.}}
\end{center}
For calculation of the cross section for pp $\rightarrow$ ALP we work in impact parameter space. The total diffractive cross section with elastic component, SD - the single dissociation and DD - diffractive dissociation components, has a  form
\[\sigma_{el+SD+DD}=\int d^2b\sum\limits_{i,k}|a_i|^2|a_k|^2|(1-e^{-\Omega_{ik}(b)/2})|^2,\]
with $\Omega_i (\Omega_k)$ - the opacity of the state i(k) probed by the incoming parton in the subprocess. This formula will be used in Monte Carlo modeling for further calculations in the paper. 

\section{Results of calculations}
As part of experimental measurements by the ATLAS collaboration during 2015 and 2018 years of light-on-light scattering during Pb+Pb collisions, at an integrated luminosity of 2.2 nb$^{-1}$ and energy of 5.02 TeV, taking into account the obtained experimental limitations \cite{5.}, we modeled the cross sections of the ALP production, candidates for dark matter, with subsequent decay into gamma quanta. Light-on-light scattering candidates are selected in two-photon events, with pseudorapidity $|\eta| < 2.37$, two-photon invariant mass $m_{\gamma\gamma}$ = 5-30 GeV, and coupling constant of the order of 10$^{-3}$ GeV$^{-1}$ \cite{6.}. The obtained calculations showed the dependence of the cross section on the number of events with ALPs (Fig. 3).
\begin{center}
\includegraphics[width=0.45\textwidth]{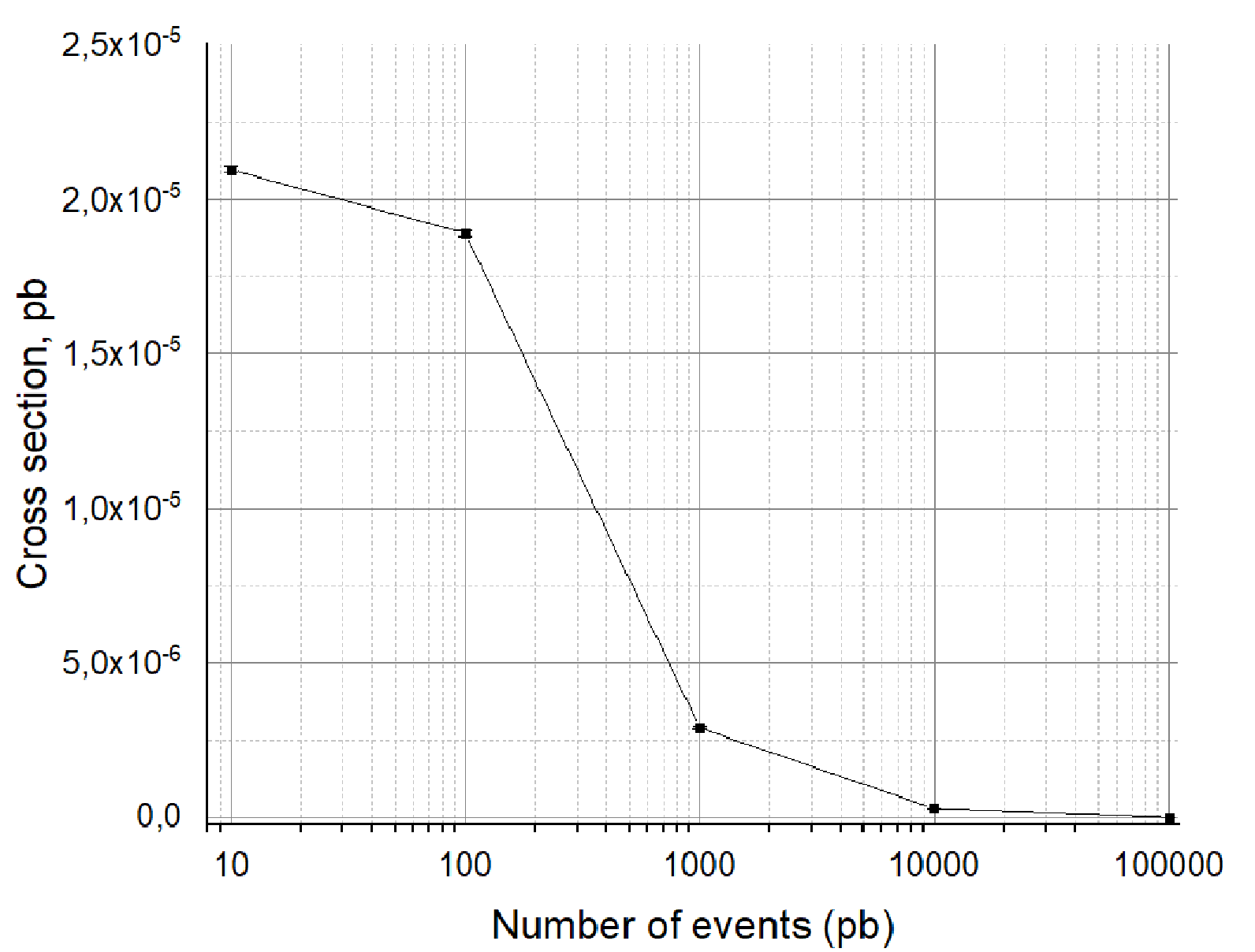}\\
\emph{\textbf{Fig.3}}{\emph{The dependence of the cross section for axon production on the number of events at the energy 5.02 TeV.}}
\end{center}
Proton-proton and ion-ion collisions were processed for QCD-initiated ALP formation using the SuperChic v.4.2 Monte Carlo event generator \cite{4.}. The corresponding cross sections were obtained and compared with recent measurements of ATLAS and CMS collaborations of light-on-light scattering at the LHC during lead-to-lead collisions \cite{7.,8.}. The cross section of ALP production in proton-proton and Pb-Pb collisions on the energy in the center of mass is presented in Fig. 4. 
\begin{center}
\includegraphics[width=0.45\textwidth]{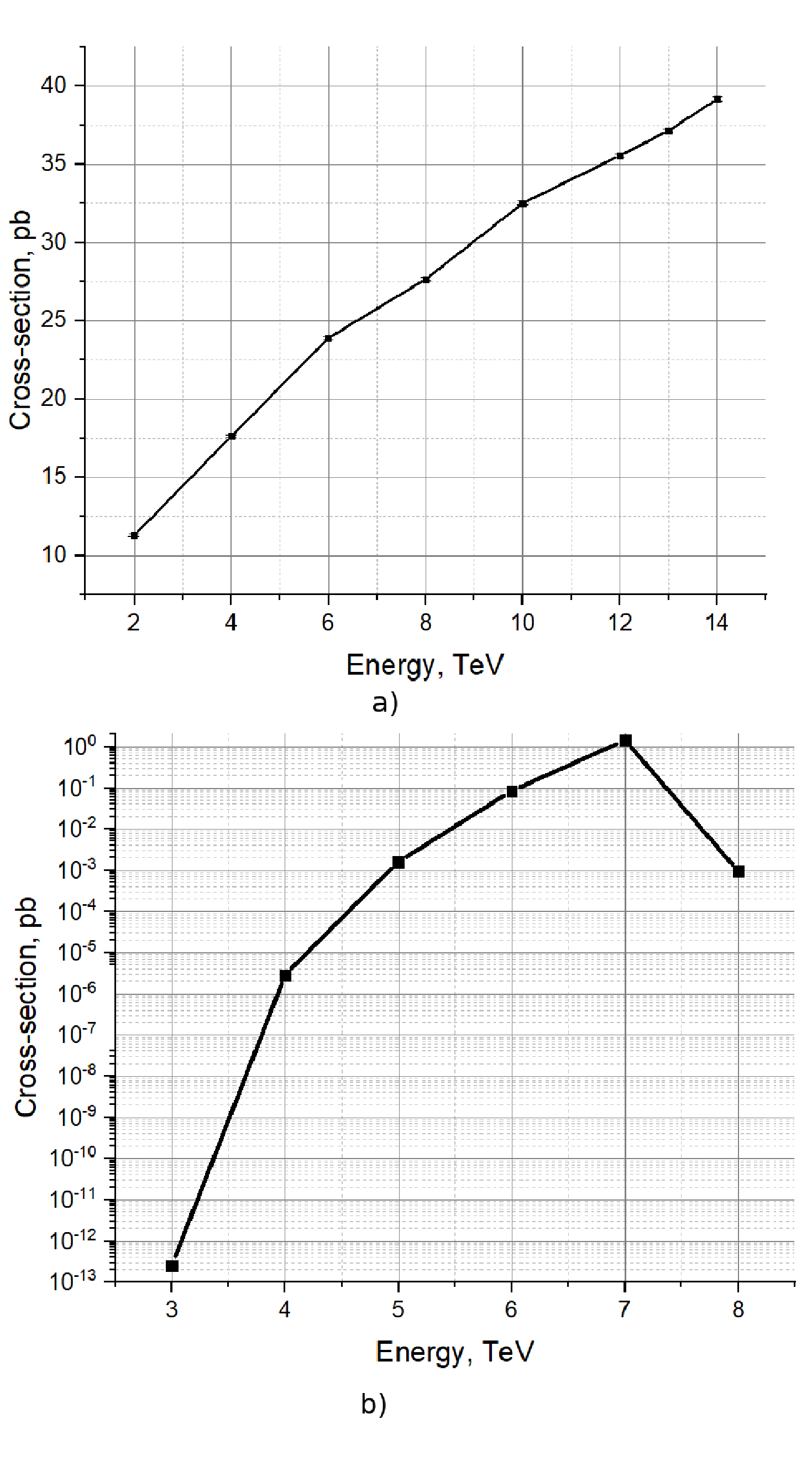}\\
\emph{\textbf{Fig.4}}{\emph{The dependence of the cross section for ALP production on energy in up: proton-proton collisions; down: Pb-Pb collisions.}}
\end{center}

Comparison of the data shown in Fig. 4 shows the different behavior of the cross section for proton-proton and Pb-Pb collisions. If for the first case we observe the increasing in the cross section with increasing proton collision energy, for lead-lead collisions the increase in the cross section from 3-7 GeV in the center of mass system is replaced by a decrease at an energy of 7-8 GeV. This indicates the multiplicity of the process associated with the formation of quark-gluon plasma and, as a result, the decrease in the cross section for ALP formation for lead-lead collisions. Using four models with the  parameters presented in Table 1, we calculated the ALP production cross sections with 1000 ALPs events in proton-proton collisions for the single dissociation and diffractive dissociation components, presented in Table 2 (1,2 models) and Table 3 (3,4 models) in the ALP mass region 5 - 30 GeV (up of the Table 2 and Table 3) and ALP mass region 5-1400 GeV (down of the Table 2 and Table 3). \\
\begin{center}
Table 2 {\it Cross sections of ALPs production for energy 5.02 TeV (up)\\ and 13 TeV (down), (1,2 models)}\\

\begin{tabular}{|c|c|c|c|} 
 \hline
   $\scriptscriptstyle{SD1}$ & $\scriptscriptstyle{DD1}$ & $\scriptscriptstyle{SD2}$ & $\scriptscriptstyle{DD2}$  \\
  \hline
 $\scriptscriptstyle{ 20.908 +/- 0.081}$ & $\scriptscriptstyle{9.167 +/- 0.079}$ & $\scriptscriptstyle{20.926 +/- 0.081}$ & $\scriptscriptstyle{9.219 +/- 0.079}$ \\
  \hline
  $\scriptscriptstyle{0.461\times 10^{-4} \pm} $ & $\scriptscriptstyle{0.176\times 10^{-4} \pm} $ & $\scriptscriptstyle{0.461\times 10^{-4} \pm} $ & $\scriptscriptstyle{0.171\times 10^{-4} \pm} $  \\
 $\scriptscriptstyle{0.31\times 10^{-6}}$&$\scriptscriptstyle{ 0.18\times 10^{-6}}$&$\scriptscriptstyle{0.31\times 10^{-6}}$&$\scriptscriptstyle{0.165\times 10^{-6}}$
\\
    \hline
    
    \end{tabular}
\end{center}

\begin{center}

Table 3 {\it Cross sections of ALPs production for energy 5.02 TeV (up) \\ and 13 TeV (down), (3,4 models)}\\

\begin{tabular}{|c|c|c|c|} 
 \hline
    $\scriptscriptstyle{SD3}$ & $\scriptscriptstyle{DD3}$ & $\scriptscriptstyle{SD4}$ & $\scriptscriptstyle{DD4}$ \\
  \hline
   $\scriptscriptstyle{20.922 +/- 0.081}$ & $\scriptscriptstyle{9.161 +/- 0.079}$ & $\scriptscriptstyle{20.905 +/- 0.081}$ & $\scriptscriptstyle{9.086 +/- 0.079 }$\\
  \hline
  $\scriptscriptstyle{0.416\times 10^{-4} \pm} $ & $\scriptscriptstyle{0.173\times 10^{-4} \pm }$ & $\scriptscriptstyle{0.419\times 10^{-4} \pm} $ & $\scriptscriptstyle{0.173\times 10^{-4} \pm} $ \\
$\scriptscriptstyle{0.26\times 10^{-6}}$&$\scriptscriptstyle{0.183\times 10^{-6}}$&$\scriptscriptstyle{0.27\times 10^{-6}}$&$\scriptscriptstyle{0.183\times 10^{-6}}$
\\
    \hline
    \end{tabular}

\end{center}

	We presented the modeled calculations for two energies, 5.02 TeV and 13 TeV, and two regions of mass, 5-30 GeV and 5-1400 GeV. Two important conclusions should be drawn from the presented data. Firstly, a single dissociation is an order of magnitude larger than diffraction dissociation at 5.02 TeV as predicted by the theory; secondly, with an increase in the ALP mass region to 1400 GeV, the cross section for the production of such particles drops by six orders of magnitude with an increase in energy to 13 TeV.

\section{Conclusions}

We presented a search for ALP in proton-proton and lead-lead scattering in association with light-by-light scattering, using the latest experimental data of ATLAS collaboration. With the help of SuperChic v4.2 program we have performed the calculations of ALP production in different energy and mass region.  We have received the dependence of the cross section for ALP production on the number of events at the energy of 5.02 TeV.  The optimal number of observed particles ranges from 10 to 100 events, which is the same order of magnitude as the latest experimental data of ATLAS collaboration \cite{9.,10.}. A search was made for a resonance in the two-photon mass distribution, corresponding to ALP with mass in the range 5-1400 GeV with new coupling constant taken from experimental constraints. As there is no dispersion and the interaction to destroy the coherence of the wave functions of the incoming protons, single dissociation is an order of magnitude larger than diffraction dissociation at 5.02 TeV. The splitting of the ALP mass region into two, 5-30 GeV and 5-1400 GeV, showed that the cross section for the production of ALPs in second region drops by six orders of magnitude with an increase in energy to 13 TeV. The results of calculations of the cross sections for the  ALP production for proton-proton and Pb-Pb collisions showed its different behavior with increasing energy in the center of mass system, and if in the first case it grows, then, for lead, the cross section begins to fall in the energy region from 7 to 8 TeV.


\begin{thebibliography}{99}                                                                                                

\bibitem{Ref1.}ATLAS-CONF-2019-002. Observation of light-by-light scattering in ultraperipheral Pb+Pb collisions with the ATLAS detector.

\bibitem{2.}ATLAS Collaboration. Search for an axion-like particle with forward proton scattering in association with photon pairs at ATLAS. \textit{JHEP} \textbf{07} (2023) 234, arXiv:2304.10953v1 [hep-ex]

\bibitem{3.} Roderik Bruce et al. New physics searches with heavy-ion collisions at the LHC. \textit{Journal of Physics G: Nuclear and Particle Physics} \textbf{47} (2020) 060501, arXiv:1812.07688v2 [hep-ph]

\bibitem{4.}Lucian Harland-Lang. SuperChic v4.2 A Monte Carlo for Central Exclusive Production. URL: https://superchic.hepforge.org/superchic4.2.pdf.

\bibitem{5.} The ATLAS collaboration. Measurement of light-by-light scattering and search for ALP with 2.2 nb?1 of Pb+Pb data with the ATLAS detector.  \textit{JHEP} \textbf{03} (2021) 243. 

\bibitem{6.}Klaudia Maj. On behalf of the ATLAS Collaboration. BSM physics using photon-photon fusion processes in UPC in Pb+Pb collisions with the ATLAS detector. ATL-PHYS-PROC-2023-037, arXiv:2307.07481v1 [hep-ex].

\bibitem{7.}ATLAS collaboration. Evidence for light-by-light scattering in heavy-ion collisions with the ATLAS detector at the LHC.\textit{ Nature Phys.} \textbf{13} (2017) 852, [arXiv:1702.01625].

\bibitem{8.}CMS collaboration. Evidence for light-by-light scattering and searches for ALPs in ultraperipheral PbPb collisions at $\sqrt{s_{NN}}$ = 5.02 TeV. \textit{Phys. Lett. B} \textbf{797} (2019) 134826, [arXiv:1810.04602].

\bibitem{9.} The ATLAS Collaboration. Search for dark matter produced in association with a Higgs boson decaying to tau leptons at \textit{s} = 13 TeV with the ATLAS detector, arXiv:2305.12938v1 [hep-ex].

\bibitem{10.} The ATLAS Collaboration. Search for ALP with forward proton scattering in association with photon pairs at ATLAS, arXiv:2304.10953v1 [hep-ex].



\end{thebibliography}
\end{document}